\documentclass{sigchi}
\toappear{Permission to make digital or hard copies of all or part of this work for personal or classroom use is      granted without fee provided that copies are not made or distributed for profit or commercial advantage and that copies bear this notice and the full citation on the first page. Copyrights for components of this work owned by others than ACM must be honored. Abstracting with credit is permitted. To copy otherwise, or republish, to post on servers or to redistribute to lists, requires prior specific permission and/or a fee. Request permissions from permissions@acm.org. \\

{\emph{CHI 2018}}, April 21--26, 2018, Montreal, QC, Canada. \\
\copyright~2018
Copyright is held by the owner/author(s). Publication rights licensed to ACM.
ACM 978-1-4503-5620-6/18/04...\$15.00. \\
https://doi.org/10.1145/3173574.3173758}
\usepackage{balance}  % to better equalize the last page
\usepackage{graphics} % for EPS, load graphicx instead
\usepackage[usenames,dvipsnames,table]{xcolor}
\usepackage[utf8]{inputenc}
\usepackage{txfonts}
\usepackage{mathptmx}
\usepackage[pdftex]{hyperref}
\usepackage{booktabs}
\usepackage{textcomp}
\usepackage{dsfont}
\usepackage{mathtools}
\usepackage{microtype} % Improved Tracking and Kerning
\usepackage{ccicons}  % Cite your images correctly!
\usepackage{multirow}
\usepackage{hhline}

\usepackage{todonotes}
\usepackage{paralist}

\usepackage{textcase}

\def\plaintitle{AdaM: Adapting Multi-User Interfaces for Collaborative Environments in Real-Time}
\def\plainauthor{
	Seonwook Park,
	Christoph Gebhardt,
	Roman R\"adle,
	Anna Feit,
	Hana Vrzakova,
	Niraj Dayama,
	Hui-Shyong Yeo,
	Clemens Klokmose,
	Aaron Quigley,
	Antti Oulasvirta,
	Otmar Hilliges
}
\def\plainkeywords{Authors' choice; of terms; separated; by
  semicolons; include commas, within terms only; required.}

\makeatletter
\def\url@leostyle{%
  \@ifundefined{selectfont}{
    \def\UrlFont{\sf}
  }{
    \def\UrlFont{\small\bf\ttfamily}
  }}
\makeatother
\def\pprw{8.5in}
\def\pprh{11in}

\definecolor{linkColor}{RGB}{6,125,233}
\hypersetup{%
  pdftitle={\plaintitle},
  pdfauthor={\plainauthor},
  pdfkeywords={\plainkeywords},
  bookmarksnumbered,
  pdfstartview={FitH},
  colorlinks,
  citecolor=black,
  filecolor=black,
  linkcolor=black,
  urlcolor=linkColor,
  breaklinks=true,
}

\begin{document}

\title{\plaintitle}

\author{
    \fontseries{b}\fontsize{12}{12}\selectfont
    Seonwook Park$^1$,
    \,
    Christoph Gebhardt$^{1}\footnotemark[2]$,
    \,
    Roman R\"adle$^{2}\footnotemark[2]$,
    \,
    Anna Maria Feit$^3$,
    \,
    Hana Vrzakova$^4$,
    \\[0.5mm]
    \fontseries{b}\fontsize{12}{12}\selectfont
    Niraj Ramesh Dayama$^3$,
    \,
    Hui-Shyong Yeo$^5$,
    \,
    Clemens N. Klokmose$^2$,
    \,
    Aaron Quigley$^5$,
    \\[0.5mm]
    \fontseries{b}\fontsize{12}{12}\selectfont
    Antti Oulasvirta$^3$,
    \,
    Otmar Hilliges$^1$
    \\[0.5mm]
    \fontseries{m}\fontsize{10.2}{12}\selectfont
    $^1$ETH Zurich
    \,
    $^2$Aarhus University
    \,
    $^3$Aalto University
    \,
    $^4$University of Eastern Finland
    \,
    $^5$University of St Andrews
}
% \numberofauthors{3}
% \author{%
%   \alignauthor{Leave Authors Anonymous\\
%     \affaddr{for Submission}\\
%     \affaddr{City, Country}\\
%     \email{e-mail address}}\\
%   \alignauthor{Leave Authors Anonymous\\
%     \affaddr{for Submission}\\
%     \affaddr{City, Country}\\
%     \email{e-mail address}}\\
%   \alignauthor{Leave Authors Anonymous\\
%     \affaddr{for Submission}\\
%     \affaddr{City, Country}\\
%     \email{e-mail address}}\\
% }

\teaser{
\centering{
	\includegraphics[width=1.0\linewidth]{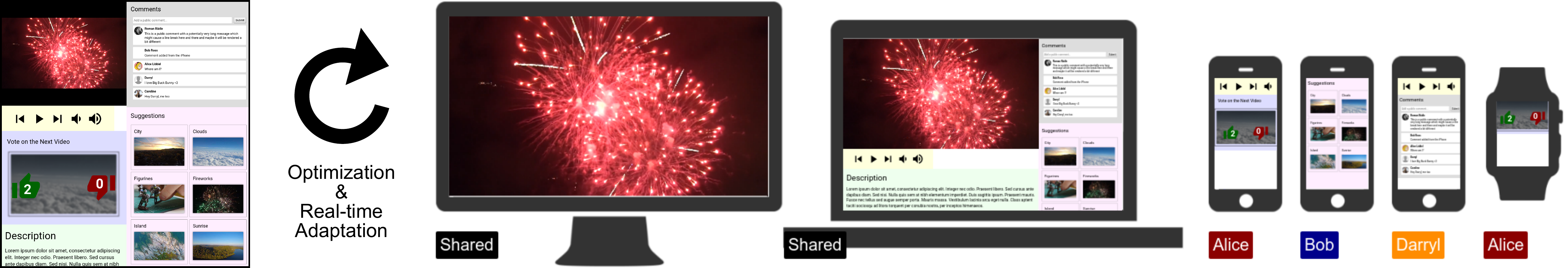}
\caption{%
Given a graphical user interface (left), AdaM automatically decides which UI elements should be displayed on each device in real-time.
Our optimization is designed for multi-user scenarios and considers user roles and preferences, device access restrictions and device characteristics.
%
%The instant adaptation also allows for designers to fine-tune the required input parameters quickly by exploring a large variety of DUI configurations .
}
\label{fig:teaser}
}
}

\maketitle

%!TEX root = ../proceedings.tex

\begin{abstract}
Developing cross-device multi-user interfaces (UIs) is a challenging problem.
There are numerous ways in which content and interactivity can be distributed. However, good solutions must consider multiple users, their roles, their preferences and access rights, as well as device capabilities.
Manual and rule-based solutions are tedious to create and do not scale to larger problems nor do they adapt to dynamic changes, such as users leaving or joining an activity.
In this paper, we cast the problem of UI distribution as an assignment problem and propose to solve it using combinatorial optimization.
We present a mixed integer programming formulation which allows real-time applications in dynamically changing collaborative settings.
It optimizes the allocation of UI elements based on device capabilities, user roles, preferences, and access rights.
We present a proof-of-concept designer-in-the-loop tool, allowing for quick solution exploration.
Finally, we compare our approach to traditional paper prototyping in a lab study.
\end{abstract}

\renewcommand{\thefootnote}{\fnsymbol{footnote}}
\footnotetext[2]{\vspace*{-1.5mm}These authors contributed equally to this work.}

\category{H.5.m.}{Information Interfaces and Presentation
  (e.g. HCI)}{Miscellaneous}


%\input{content/01_introduction}
%!TEX root = ../proceedings.tex

\section{Introduction}
Many users now carry not one, but several computing devices, such as laptops, smartphones or wearable devices. In addition, our environments are often populated with public and semi-public displays.
In collaborative settings, such as at work or in education, many application scenarios could benefit from UIs that are distributed across available devices and potentially also across multiple users participating in a joint activity.
However, traditional interfaces are designed for a single device and are neither aware nor do they benefit from having multiple input and output channels available.
This may be ascribed, in part, to the significant complexity of designing and implementing such cross device interfaces and the combinatorial complexity of the question of which UI element should be placed onto which of the users' devices.

Our goal is to provide computational support for the task of distributing elements in a rapid and controllable way among devices in a collaborative setting.
Consider a concert, exhibition, birthday party, or a work meeting:
depending on their device capabilities, co-present users would have parts of an interface displayed on their devices.
Instead of device owners manually deciding \emph{assignment} (who gets what),
elements are automatically distributed such that the most important elements are always available while taking into account personal preferences and constraints including privacy.
Such collaborative settings are inherently dynamic with users and devices appearing and disappearing at various points in time.
This requires a real-time approach to accommodate dynamic device configurations, user preferences and user roles.

Prior work on cross-device interfaces have proposed methods for synchronizing elements across devices \cite{Frosini:2014:UID:2607023.2607032,Klokmose:2015:WSD:2807442.2807446,Nebeling:2017:XSG:3025453.3025547,Nebeling:2016:XUC:2858036.2858048,Yang:2014:PEE:2556288.2557199} or distributing elements of a workspace over multiple displays \cite{Radle:2014:HSM:2669485.2669500,Schreiner:2015:CFC:2702613.2732909}.
Panelrama \cite{Yang:2014:PEE:2556288.2557199} uses a suitability measure for associating (single user) UI panels to devices with an integer programming formulation.
Frosini and Patern\`{o} \cite{Frosini:2014:UID:2607023.2607032} present a conceptual framework which considers multi-user roles but does not provide methods to solve the assignment problem.
Prior to this paper, no automatic solutions existed for element distribution in collaborative settings which considers critical constraints such as access rights, privacy, and roles and their dynamic evolution over time.

We propose an optimization-based approach that automatically distributes elements to available devices by solving a many-to-many assignment problem, constraining the optimization by available screen real-estate.
Given a list of UI elements
and available devices, user and device descriptions, it distributes the UI elements based on an objective that maximizes the usefulness of an element on a device while simultaneously maximizing completeness of the UI from a user's perspective (i.e., ensuring that important elements are present for each user).
More precisely our method
\begin{inparaenum}[(1)]
 \item takes role requirements and
 \item user preferences into account when distributing elements,
 \item adapts to changing user roles or preferences depending on a given task,
and
 \item adapts the DUI in real-time based on presence of users and devices in collaborative scenarios.
\end{inparaenum}
Our formulation can be solved quickly, easily scaling up to thousands of users and devices.
The benefit to users and designers is the new type of control provided: instead of instructing \emph{how} elements should be distributed (a heuristic or rule-based approach),
or completing it manually,
developers and designers can express qualities of ``good'' distributions.
As shown in Figure~\ref{fig:teaser} this control offers substantial promise for the creation of applications that effectively take advantage of the wide range of capabilities in cross-device ecosystems for collaborative multi-user interfaces.

We demonstrate the utility of our approach with a step-by-step walkthrough of how the system adapts to various roles and preferences in a company meeting setting, and demonstrate real-time adaptiveness in a fully implemented co-located media sharing application.
Furthermore, we suggest how the algorithm could scale to address previously impossible problem scales.
In addition, we evaluate our approach in a user study and compare it to traditional paper prototyping.

%!TEX root = ../proceedings.tex

\section{Related Work}

Cross-device or ``Distributed User Interfaces'' (DUIs) offer appealing features including, more pixels~\cite{Schwarz:2012:PPE:2207676.2208378}, new forms of engagement at varying scales~\cite{Terrenghi:2009:TAM:1644246.1644265}, reduction in system complexity by splitting and sharing functionality~\cite{Dong:2016:UCD:2901790.2901851} and targeting interactions across and between devices (e.g., \cite{Nebeling:2014:IDC:2611247.2556980}). This vision has given rise to sustained research interest within the HCI community from research on taxonomies~\cite{Terrenghi:2009:TAM:1644246.1644265,Nebeling:2016:CIE:2992154.2996361}, interaction techniques, and middleware~\cite{Klokmose:2015:WSD:2807442.2807446}.
We briefly discuss related work across several related areas from DUIs to UI optimization.

\subsection{Cross Device User Interfaces}
People now use multiple devices with displays (e.g., laptops, phones, tablets), often at the same time. Commercial software solutions exist for mirroring (e.g., AirPlay), I/O targeting (e.g., Microsoft Continuum), coordinating (e.g., Apple Continuity) or stitching multiple displays (e.g., Equalizer~\cite{Eilemann:2009:Equalizer}) into a single canvas. However, design and development for such settings is entirely manual and requires the developer to consider the myriad set of inputs, outputs and device configurations to achieve even rudimentary cross-device experiences. When designing for multiple users this problem is further exacerbated due to access rights, privacy and user preference concerns.

Existing cross-device research has highlighted challenges in adapting DUIs for collaborative environments in real-time, including problems in testing multi-device experiences~\cite{Dong:2016:UCD:2901790.2901851}, user interface widget adoption~\cite{Grubert2016}, functional UI coordination ~\cite{Santosa:2013:FSM:2493432.2493476}, component role allocation~\cite{Waljas:2010:CSU:1851600.1851637}, spatial awareness~\cite{Radle:2015:SSE:2702123.2702287} and changes in related parallel use~\cite{Jokela:2015:DSC:2702123.2702211}. Addressing these challenges has given rise to the approach taken here.

Our work is concerned with computational support for the design of distributed or cross-device UIs \cite{Elmqvist2011,Nebeling:2016:CIE:2992154.2996361} in the sense of a \emph{crossmedia service} where the functionality of a single application is decomposed and shared across devices and users. We propose an algorithmic approach to functionality assignment according to device strengths and user preferences, extending prior rule-based approaches \cite{Husmann:2017:OMP:3078810.3078812,Nebeling:2017:XSG:3025453.3025547}. Functionality distribution to different devices is a crucial element of DUI design since a balanced assignment of interactive components can reduce the complexity of the original system \cite{Dong:2016:UCD:2901790.2901851}.

Rule-based approaches \cite{Husmann:2017:OMP:3078810.3078812,Nebeling:2017:XSG:3025453.3025547} provide insights into cross-device interaction patterns in the real-world but do not scale to many devices or multi-user scenarios.
We believe that our bottom-up approach of modeling DUI usability in multi-user scenarios opens up unexplored application areas.

\subsection{Toolkits and Middleware}
Existing toolkits have explored cross device interaction with combinations of mobile devices \cite{Hamilton:2014:CEU:2556288.2557170, O'Leary:2017:MCK:3064663.3064768, Radle:2014:HSM:2669485.2669500, Schreiner:2015:CFC:2702613.2732909, Schwarz:2012:PPE:2207676.2208378}, mobile/desktop devices \cite{Heikkinen:2014:TBT:2611009.2611026, Melchior:2009:TPD:1570433.1570449, Nebeling:2017:XSG:3025453.3025547, Nebeling:2014:IDC:2611247.2556980}, mobile/display wall devices \cite{Badam:2014:PCF:2669485.2669518, Klokmose:2015:WSD:2807442.2807446} and wearables \cite{Chi:2015:WSC:2702123.2702451, Grubert:2015:MMF:2702123.2702331, Houben:2015:WTP:2702123.2702215}. Alternative approaches have focused on the development of conceptual frameworks \cite{vistiles2017, Paterno:2012:LFM:2305484.2305494}. Within this work, common applications which support multiple people, with cross device interactions, include authoring~\cite{Klokmose:2015:WSD:2807442.2807446}, web browsing \cite{Ghiani:2010:OCI:1851600.1851653,Han:2000:WUX:358916.358993,luyten:2005:dui} and collaborative visualizations \cite{Badam:2014:PCF:2669485.2669518}.

Prior work has often focused on providing support for keeping application and UI states synced across devices using conventional software development practices~\cite{Klokmose:2015:WSD:2807442.2807446}.
Our work builds on these capabilities to go beyond the state-of-the art in the automatic distribution of UI elements to users and devices.

\subsection{Mobile Co-located Interaction and Collaboration}
DUIs have emerged as a platform of interest for supporting mobile co-located interaction~\cite{Lucero:2013:MCI:2427076.2427083}. Existing research has investigated systems that allow groups of co-located people to collaborate around a digital whiteboard with mobile devices (e.g., PDAs) \cite{hilliges2007designing,Marquardt:2012:CIV:2380116.2380121,Myers:1998:CUM:289444.289503,Rekimoto:1998:MDA:274644.274692}. With mobile devices alone, research has explored co-located collaboration for shopping \cite{Robinson:2017:BTD:3098279.3098534}, video \cite{Robinson:2017:BTD:3098279.3098534}, ideation \cite{Porcheron:2016:CDM:2994310.2994350} and content sharing  \cite{Lucero:2011:PCU:1978942.1979201,Marquardt:2012:CIV:2380116.2380121}. Our work explicitly targets heterogeneous settings where devices with different capabilities are used to create a single collaborative system. By considering each user separately, we also allow for better distribution of functionality across homogeneous devices. This can occur often with mobile phones in mobile co-located interactions. Additionally, we address the dynamicity of mobile interactions in terms of available users and devices by providing a real-time formulation. %%tofix

\subsection{Computational UI Generation and Retargeting}
Modern optimization methods have been proposed to automate UI generation and retargeting. SUPPLE \cite{Gajos:2004:SAG:964442.964461} uses decision-theoretic optimization to automatically generate UIs adapted to a person's abilities and computational solutions have been shown for example in PUC \cite{Nichols:2002:GRC:571985.572008}, automatically creating control interfaces for complex appliances.
Smart Templates \cite{Nichols:2004:IAI:964442.964507} uses parameterized templates to specify when to automatically apply standard design conventions.
One important observation that we build on in this work is that many GUI design problems such as layout of menus, web pages, and keyboards can be formulated as an assignment problem \cite{karrenbauer2014improvements,oulasvirta2017user}.

Model-based approaches for UI retargeting have proposed formal abstractions of user interfaces (UIDLs) to describe the interface and its properties, operation logic, and relationships to other parts of the system \cite{eisenstein2001applying} which can then be used to compile interfaces in different languages and to port them across platforms.
Data-driven approaches have been explored by Kulkarni and Klemmer \cite{Kulkarni:2011:AAW:1979742.1979810} to automatically transform desktop-optimized pages to other devices.
GUMMY \cite{Meskens:2008:GMU:1385569.1385607} retargets UIs from one platform to another by adapting and combining features of the original UI.

To the best of our knowledge, no prior work addresses the computational assignment of UI elements to devices in multi-user settings that would consider critical constraints such as access rights, privacy, and roles and their dynamic evolution over time.
AdaM provides a real-time capable optimization formulation and implementation using mixed integer linear programming. %%tofix

%!TEX root = ../proceedings.tex

\section{Concepts}

The type of scenarios we consider in this work are co-located multi-user events -- such as a meeting, party, or lecture. Any number of people with various devices and roles can be involved.
An interactive application is assumed to consist of elements of different types,
and the participants show varying interest toward them, but not all devices can show all elements.
We further assume that this setup and the need for interactivity can change dynamically as time progresses.
In order to cast such scenarios for combinatorial optimization,
we need to introduce and define a few central concepts.
These concepts are the basis for the objectives and constraints of the assignment problem formulation we develop in the next section.

\paragraph{Element Importance}
Depending on the preferences of users present, the display of some elements should be prioritized.
For example, in the lecture scenario the slides need to be presented on a public display, whereas a chat channel for the audience may only be displayed if auxiliary, personal devices (e.g., phones) are available.
This importance value may be defined by the application developer or user.
Element importance is one of the aspects an optimization scheme needs to consider and trade-off with other, potentially contradictory, preferences.

\paragraph{Device Access}
In collaborative settings we assume that personal devices as well as shared devices must be considered.
An example of a shared device is a large screen in a conference room, whereas a private device can range from smart wearables to laptop computers.
In order to apply a user's preferences through the importance metric, we must know which devices are available to a user.
Thus, we can describe the user's access to a device by its availability to the user, defined either in terms of ownership or physical proximity.

\paragraph{Element Permission}
We integrate user roles into our optimization scheme by considering that some elements should not be made available to specific users.
For example, while a disc jockey may require access to the audio mixer UI, the light technician should focus on stage lighting and the stage crew should not have access to either.
To effectively represent such user roles in the final DUI, one would have to make sure that users are authenticated properly.
We assume that mechanisms for this exist.

\paragraph{Device Characteristics}
An element which requires frequent and quick text input should be assigned to a device with either a physical or soft keyboard (e.g., laptop, phone)  rather than to a display only (e.g., TV).
Similarly, visually rich elements such as presentation slides or a video should not be placed on small-screen devices (e.g., smartwatch).
Similar to Panelrama \cite{Yang:2014:PEE:2556288.2557199}, we consider visual quality, pointing and text input mechanisms as device characteristics.

\paragraph{Element Requirements}
Complementary to device requirements we also define element requirements.
Not all elements can be shown on every terminal.
An element such as a drawing canvas may require precise pointing input as well as high visual fidelity, where assignment to a touchscreen tablet would be preferred over a small phone.

%!TEX root = ../proceedings.tex

\section{Optimization Formulation}

\newcommand{\E}[0]{\ensuremath\mathds{E}}
\newcommand{\D}[0]{\ensuremath\mathds{D}}
\newcommand{\U}[0]{\ensuremath\mathds{U}}

With above concepts in place we develop a formalization as a mixed integer program that can be solved with state-of-the-art ILP solvers such as Gurobi \cite{gurobi}.
These solvers can automatically search for solutions that maximize the objective and satisfy the defined constraints while assigning integer solutions to decision variables and give formal bounds on the solution quality with respect to the objective function.
In the following, we define the overall objective. The subsections define each of its terms in detail.

\subsection{Main Objective and Decisions}
To begin, we identify that device access, element permission, and element privacy are concepts which constrain our problem.
On the other hand, element importance, device characteristics, and element requirements directly address our objective of building a usable DUI.
We thus propose a conceptually simple objective with the sub-objectives of: quality ($Q$) and completeness ($C$), which we aim to maximize in our final assignments.
Here $Q$ measures whether the correct elements are assigned to a user and device and $C$ measures whether a user receives all necessary elements.
We formulate our objective function as a weighted sum of the normalized terms ($\in [0,1]$):
\begin{equation}
	\max_{e,\,d}~
	w_q \hat{Q}
	+ w_c \hat{C},
	\label{eq:objective}
\end{equation}
where $w_q + w_c = 1$.
We empirically set $w_q = 0.8$.
Elements $e\in\E$, devices $d\in\D$, and users $u\in\U$ are considered.

In this study, we only consider the assignment of elements to devices.
The problem of layout of elements on a device is assumed to be performed by responsive design practices common in web design. In our Demo Application section we demonstrate how a thin layer of UI code is sufficient to create fully functional user facing applications.

At the core of our method lies the decision on how to assign element $e$ to device $d$, defined as,
\begin{equation}
\begin{split}
	x_{ed} =
	\left\{
	\begin{matrix*}[l]
		1 & \mathrm{if}~e~\mathrm{assigned~to}~d \\
		0 & \mathrm{otherwise}
	\end{matrix*}
	\right.
\end{split}
\end{equation}
All other decision variables pertaining to secondary optimization criteria such as element size and element count (per user) and input parameters are defined in Table~\ref{tab:optimization-inputs}.

\begin{table}
	\uppercase{\scriptsize Decision Variables}\\[0.1cm]\normalsize
	\def\arraystretch{1.3}
	\begin{tabular}{|>{\raggedleft}m{4.5mm}@{\hspace{0.5mm}}>{\raggedright}m{11.8mm}|m{63mm}|}
		\hline
		\multicolumn{2}{|c|}{Variable} & Description \\
		\hline
		$x_{ed}$ & $\in\{0,\,1\}$ & Assignment of element $e$ to device $d$ \\
		$s_{ed}$ & $\in\mathbb{Z}^+$ & Area of element $e$ on device $d$ \\
		$o_{eu}$ & $\in\{0,\,1\}$ & Whether element $e$ is made available to user $u$ \\
		$r_\mathrm{min}$ & $\in\mathbb{R}^+_0$ & Minimum elements-coverage over all users \\
		\hline
	\end{tabular}\\[0.1cm]

	\uppercase{\scriptsize Input Parameters}\\[0.1cm]\normalsize
	\begin{tabular}{|>{\raggedleft}m{4.5mm}@{\hspace{0.5mm}}>{\raggedright}m{11.8mm}|m{63mm}|}
		\hline
		\multicolumn{2}{|c|}{Parameter} & Description \\
		\hline
		$a_{ud}$ & $\in\{0,\,1\}$ & Whether user $u$ has access to device $d$ \\
		$p_{eu}$ & $\in\{0,\,1\}$ & Whether user $u$ is given permission to interact with element $e$ \\
		$i_{eu}$ & $\in[0,\,1]$   & Importance of element $e$ to user $u$ \\
		$\mathbf{u}_d$ & $\in[0,\,1]$ & Device characteristics vector \\
		$\mathbf{v}_e$ & $\in[0,\,1]$ & Element requirements vector \\
		\hline
	\end{tabular}\\[-0.5mm]
	\begin{tabular}{|>{\centering}m{16.8mm}|m{63mm}|}
		\hline
		$w_e^\mathrm{min}\times h_e^\mathrm{min}$ & Minimum size of element $e$ in pixels \\
		$w_e^\mathrm{max}\times h_e^\mathrm{max}$ & Maximum size of element $e$ in pixels \\
		$w_d\times h_d$ & Size of screen on device $d$ in pixels \\
		\hline
	\end{tabular}\\[0mm]
	\caption{Description and ranges of variables and input parameters.}
	\label{tab:optimization-inputs}
\end{table}

\subsection{Quality Term ($Q$)}
The quality of the final assignment relies on the suitability of assigning an element $e$ to device $d$ in terms of device characteristics $\mathbf{u}_d$ and element requirements $\mathbf{v}_e$.
$\mathbf{u}_d$ and $\mathbf{v}_e$ are $4$-element vectors with values in range $[0,\,1]$.
The values represent visual quality and availability of text input, touch pointing, and mouse pointing.
This is similar to the approach in \cite{Yang:2014:PEE:2556288.2557199}.

In addition, we take users' preferences through $i_{eu}$ into account and consider the area that an element would occupy on a device.
As an element cannot take up more space than is available on the display of a device, this consideration proves to be crucial for ensuring that not all elements are assigned to every device.
For each device, a mean importance $i_{ed}$ is calculated over all users who have access to this device.
By taking the mean, we aim to balance the preferences of multiple users.
We also aim to maximize the size of more important and compatible elements.
That is, elements which are capable of being larger and benefit from additional screen real-estate (e.g., HD video) should be allowed to do so.
Hence, we assume that a larger version of an element exhibits better visual quality than a smaller version.

The final quality term is then defined as:
\begin{equation}
	Q = \sum_e \sum_d c_{ed} i_{ed} s_{ed}
	\label{eq:quality_term}
\end{equation}
\vskip -4mm
where,
\begin{equation*}
	c_{ed} = \mathbf{u}_d \cdot \mathbf{v}_e
	\quad \mathrm{and} \quad
	i_{ed} = \sum_u i_{eu} a_{ud} / \sum_u a_{ud}
\end{equation*}
are combined input parameters describing device and element characteristics ($c_{ed}$) and importance of element to user ($i_{ed}$).

\subsection{Completeness Term ($C$)}
When assigning elements across devices, we must furthermore consider and ensure the usefulness of the resulting UI from each user's perspective.
With the element permission parameter $p_{eu}$, we define a subset of elements which a user should be able to interact with.
To ensure that the DUI is complete in the sense that all necessary functionality can be accessed by a given user in a collaborative multi-user scenario, we explicitly model the completeness of the UI per user.

Intuitively the completeness of the DUI for a user can be defined by:
\vspace*{-2mm}
\begin{equation}
	r_u = \frac{\sum_e o_{eu}}{\sum_e p_{eu}} \qquad\forall e,\,u
\end{equation}
where,
\begin{equation}
	o_{eu} = \begin{cases}
		1 & \textrm{if~}\sum_d a_{ud} x_{ed} > 0 \\
		0 & \textrm{otherwise}
	\end{cases}
	\qquad \forall e,\,u
	\label{eq:comp-max}
\end{equation}

The completeness variable $r_u$ describes the proportion of UI elements that a user has access to.
A user with $r_u=1$ would have access to all elements which she requires for her role, that is, all elements with $p_{eu}=1$.

The decision variable $o_{eu}$ represents whether an element $e$ has been made available by assignment to a user $u$, taking into account the devices for which the user has access to (i.e., where $a_{ud}=1$).
This variable is determined by maximizing our objective \eqref{eq:objective} and applying the following constraints:
\begin{equation}
	o_{eu} \leq 1
	\quad \mathrm{and} \quad
	o_{eu} \leq \sum_d a_{ud} x_{ed}.
\end{equation}

In addition, we consider the least privileged user, that is, the user with lowest $r_u$.
This variable is denoted $r_\mathrm{min}$ and it is determined by applying the following additional constraints:
\begin{equation}
	r_\mathrm{min} \geq 0
	\quad \mathrm{and} \quad
	r_\mathrm{min} \leq \frac{\sum_e o_{eu}}{\sum_e p_{eu}} \qquad \forall u
\end{equation}

We now formulate the completeness term in the objective as,
\begin{equation}
	C = \sum_u \sum_e o_{eu} + r_\mathrm{min}
	\label{eq:completeness_term}
\end{equation}
where we maximize the mean UI completeness of users, and in particular try to improve the result for users with $r_\mathrm{min}$ coverage.

\subsection{Assignment Constraints}
The previous terms alone cannot sufficiently constrain the optimization.
In particular, we cannot support private elements or limit the assignment of elements in a meaningful way.
In this section, we describe state and describe the constraints which allow for an effective optimization formulation.

\paragraph{Element Area Constraint}
The element size variable $s_{ed}$ must be determined based on whether an element $e$ is assigned to a device $d$ at all.
We thus define the following for all $e,\,d$:
\begin{align}
	\begin{split}
        x_{ed} = 0 &~\implies~ s_{ed} = 0  \\
        x_{ed} = 1 &~\implies~ s_e^\mathrm{min} \leq s_{ed} \leq \min\left(s_e^\mathrm{max},s_d\right)  ,
	\end{split}
\end{align}
ensuring that the area of an element be zero if it is not assigned and that it lies between user-specified bounds otherwise.

\paragraph{Device Capacity Constraint}
In Eq. \eqref{eq:quality_term}, we aim to maximize the size of all elements.
We constrain this maximization by saying that the assignment of element sizes should not exceed the device's display area.
An assumption is made to say that a sum of the area of rectangular elements $e$ assigned on device $d$ represents the total area used by the elements.
While this assumption would not always hold, it works in practice as shown in our evaluations. The device capacity constraint is formulated as follows:
\begin{align}
\begin{split}
	s_{ed} &\leq m x_{ed} \qquad \forall e,\,d \\
	\sum_{e} s_{ed} &\leq w_d h_d
	\qquad~~\, \forall d.
\end{split}
\end{align}
where $m$ is a sufficiently large number.

Due to our simplifying assumption, we must explicitly ensure that the minimal width and height of an element allows it to be assigned to a device.
This is expressed with the following constraints:
\vspace*{-4mm}
\begin{align}
\begin{split}
w_e^\mathrm{min} x_{ed} &\leq w_d
\qquad \forall e,\,d \\
h_e^\mathrm{min} x_{ed} &\leq h_d
\qquad \forall e,\,d  .
\end{split}
\end{align}

\paragraph{Element Permission Constraint}
When assigning an element, we must consider the element permissions variable $p_{eu}$, which must be evaluated for every assignment $x_{ed}$.
We do this by considering a device $d$ for which some users have access ($a_{ud} = 1$).
If any of these users do not have permission to interact with an element $e$ (i.e., $p_{eu} = 0$), then the element should not be assigned to the device.
This is expressed as:
\begin{equation}
	\sum_u \mathds{1}\left(a_{ud} > p_{eu}\right) > 0
        ~\implies~
	x_{ed} = 0 \qquad
	\forall e,\,d  .
\end{equation}

\paragraph{Device Accessibility Constraints}
Furthermore, a device which is accessible by none of the users should not have any elements assigned,
\begin{equation}
  x_{ed} \leq \sum_{u} a_{ud}
\qquad \forall e,\,d  .
\label{eq:inaccessible_constraint}
\end{equation}

\paragraph{Zero Constraints}
Finally, we check if the compatibility or importance of an assignment $x_{ed}$ is zero with:
\begin{align}
\begin{split}
	c_{ed} &= 0 ~\implies~ x_{ed} = 0 \qquad \forall e,\,d \\
	i_{ed} &= 0 ~\implies~ x_{ed} = 0 \qquad \forall e,\,d
\end{split}
\end{align}
We apply these constraints to make a distinction between very low importance or compatibility and zero-value input parameters.
This allows for users to express a definite decision against an element assignment.

\paragraph{User-defined Element Assignment}
Though not shown, our work may simply be extended to give users explicit control on element-device assignment.
For instance, to ensure that element $\tilde e$ is assigned on device $\tilde d$, the constraint $x_{\tilde e \tilde d} = 1$ could be added.
Similarly, $x_{\tilde e \tilde d}=0$ can ensure that $\tilde e$ is not assigned to $\tilde d$.
Note that the user-facing application should account for cases where the additional constraint cannot be fulfilled such as when minimum element size exceeds device capacity.

%!TEX root = ../proceedings.tex

\section{AdaM Design Tool}

The AdaM Design Tool is a proof-of-concept designer-in-the-loop tool that allows for rapid solution space exploration. It consists of the AdaM Application Prototype and the AdaM Simulator. The Application Prototype allows the designer to specify input parameters required by the optimizer to allocate elements to devices and automatically applies the optimizer result. The simulator allows for quick tuning of input parameters by applying changes in device configurations immediately.

We build our tool on top of Codestrates \cite{radle2017codestrates} and Webstrates \cite{Klokmose:2015:WSD:2807442.2807446}, which transparently synchronize the state of the Document Object Model (DOM) of webpages. Codestrates further enables collaborative prototyping and rapid iterations of AdaM applications. Communication with the optimizer back-end happens over a websocket connection.

\subsection{AdaM Application Prototype}

The AdaM Application Prototype includes an integrated development environment (IDE) for editing application content and behavior, as well as a configuration panel UI that allows for changing the parameters of optimizable elements. The platform is web-based and each AdaM application is a single web-page that contains optimizable elements, that it can hide or show based on the optimized solution.

\begin{figure}[tb]
\centering
\includegraphics[width=\columnwidth]{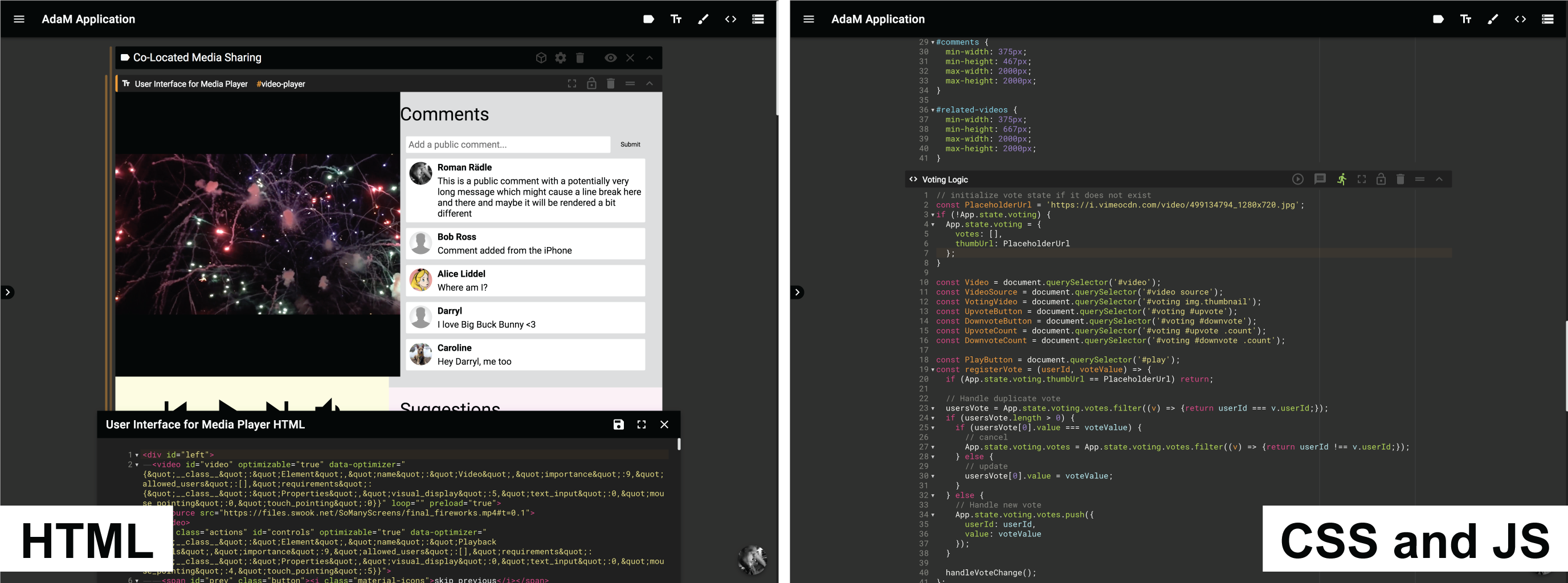}
\caption{AdaM application in edit mode with the HTML of an application (left) and its CSS and JavaScript (right).}
\label{fig:adam-platform}
\end{figure}

The designer can develop the user interface and the interactive behavior of an AdaM application using standard HTML5, JavaScript, and CSS3 (Figure~\ref{fig:adam-platform}). A final application can be put into fullscreen (Figure~\ref{fig:teaser}). All changes are instantly reflected in the browser, allowing for rapid application development and testing. Each application is addressed by a URL, which can be shared with others to collaboratively develop applications or to run it on devices.

\begin{figure}[tb]
\centering
\includegraphics[width=\columnwidth]{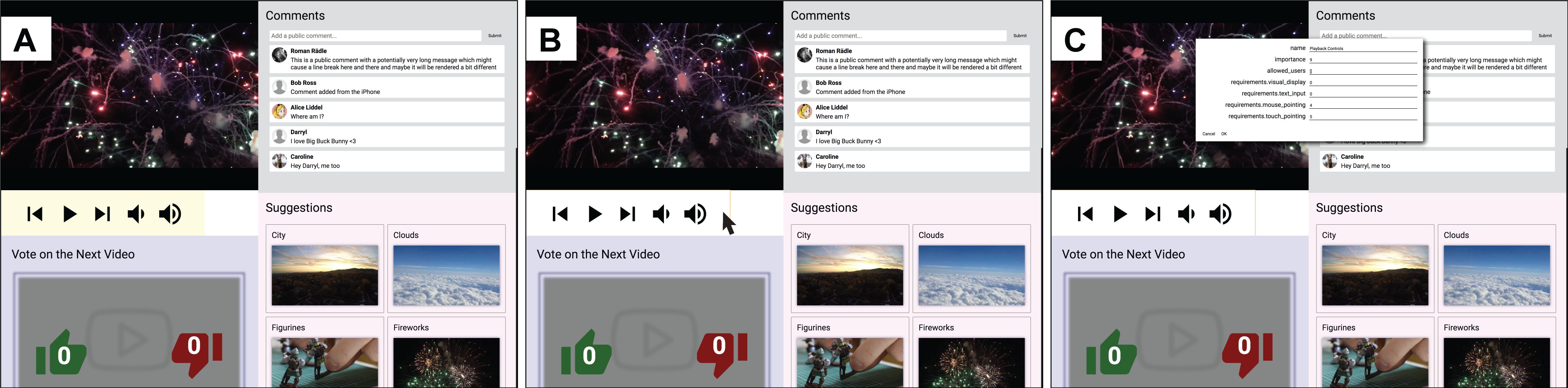}
\caption{Workflow to open configuration panel UI to specify element parameters. Media sharing application (A) with highlighted optimizable controls element (B) and open configuration panel (C).}
\label{fig:adam-configuration-panel}
\end{figure}

The designer has to annotate HTML elements with the attribute \texttt{optimizable="true"} to consider them for optimization. Initially, the optimizer uses default parameters for elements but they can be specified by the designer. Pressing the control key on the keyboard and clicking on an optimizable element opens the configuration panel UI (Figure~\ref{fig:adam-configuration-panel}). This panel allows the designer to enter parameters related to default element-user importance, element requirements, and user permissions.

Each AdaM application communicates with the optimizer back-end by sending changes of its state (e.g., when the application has loaded, or parameters for elements that have changed), and receives updates from the optimizer including updates caused by other clients. A change includes updated user-specified parameters and user/device configuration. Device information are automatically read out from the device (e.g., window width and height) or can be set as URL parameters. This is useful for testing with different devices.

\subsection{AdaM Simulator}
Testing multi-device user interfaces is inherently difficult, as it requires managing the input and output of multiple (often heterogeneous) devices at the same time.
To overcome this challenge, we developed a simulator that allows us to instantiate a wide range of simulated devices in a web browser and control the device characteristics used by the optimizer. A device is simulated in an iframe pointing to a given AdaM application, parameterized to e.g., act like a user's personal tablet or a shared interactive whiteboard.

The simulator has a pre-defined set of a device types from which the designer can choose (i.e., TV, laptop, tablet, smartphone, and smartwatch) (Figure~\ref{fig:teaser}). The simulated device characteristics can be changed at any time. For example, user access, device display dimensions, or device affordances. A device in the simulator can be disabled to simulate a device leaving or enabled to simulate a device joining.

%!TEX root = ../proceedings.tex

\definecolor{Assistant}{rgb}{0.05,0.04,0.41}
\definecolor{Presenter}{rgb}{0.1,0.32,0.09}

\definecolor{Presentation}{rgb}{0,1,0.01}
\definecolor{PresenterControls}{rgb}{0.25,0.88,0.82}
\definecolor{PresenterNotes}{rgb}{0.8,0.36,0.36}
\definecolor{Clock}{rgb}{0,0.75,1}
\definecolor{QuaterlyFigures}{rgb}{1,0.65,0}
\definecolor{MinutesView}{rgb}{1,0.41,0.7}
\definecolor{MinutesEdit}{rgb}{0.95,0.9,0.55}
\definecolor{EmployeesNotes}{rgb}{0.6,0.98,0.6}

\section{System Walkthrough}
To illustrate the utility of our approach, we start by discussing simple scenarios first, building up to more challenging scenarios and a functional end-to-end application.
The initial illustrative examples build on a meeting room scenario.
There are four users present in this scenario: the manager (`boss'), her assistant, an employee, and a colleague who is presenting work results.
We adjust specific parameters of our formulation per scenario and illustrate the effects.

\subsection{A. User Roles}
By considering user roles in our constraints, we can ensure that a particular user does not receive elements irrelevant to their role and task.
A first simple use case involves the presenter and assistant.
We set binary permission values between elements and user, defining the UI elements each role has access to (but not the assignment of elements to devices).
For the purposes of our demonstration we only consider three UI elements:

\renewcommand*{\arraystretch}{1.4}
\hspace*{\fill}
\begin{tabular}{r|>{\centering}p{1.5cm}|>{\centering}p{1.5cm}|>{\centering}p{1.5cm}|}
	\hhline{~|-|-|-|}
	& \cellcolor{PresenterControls} Presenter Controls
	& \cellcolor{MinutesView}       Minutes (View)
	& \cellcolor{MinutesEdit}       Minutes (Edit) \tabularnewline
	\hhline{~|-|-|-|}
	\textcolor{Presenter}{Presenter}
	& Yes & Yes & No  \tabularnewline
	\hhline{~|-|-|-|}
	\textcolor{Assistant}{Assistant}
	& No  & No  & Yes \tabularnewline
	\hhline{~|-|-|-|}
\end{tabular}
\hspace*{\fill}

\begin{figure}[!hb]
\centering
\begin{minipage}{0.48\columnwidth}
	\includegraphics[width=\columnwidth]{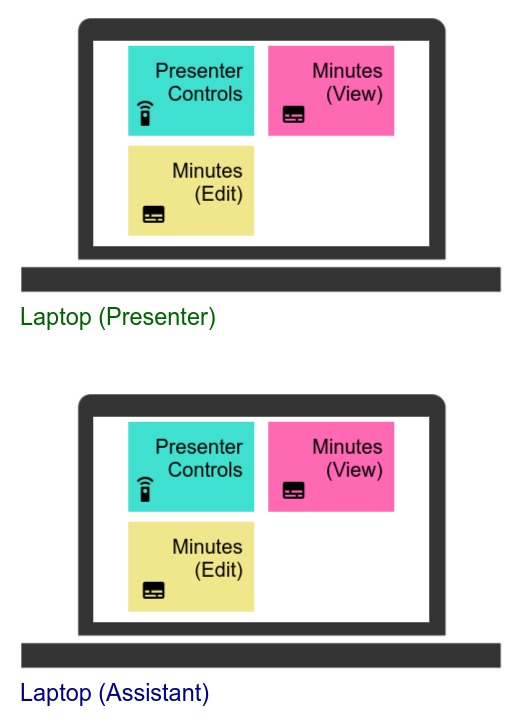}
	\vskip -3mm
	\small\raggedleft (a) Initial
\end{minipage}
\hfill\vline\hfill
\begin{minipage}{0.48\columnwidth}
	\includegraphics[width=\columnwidth]{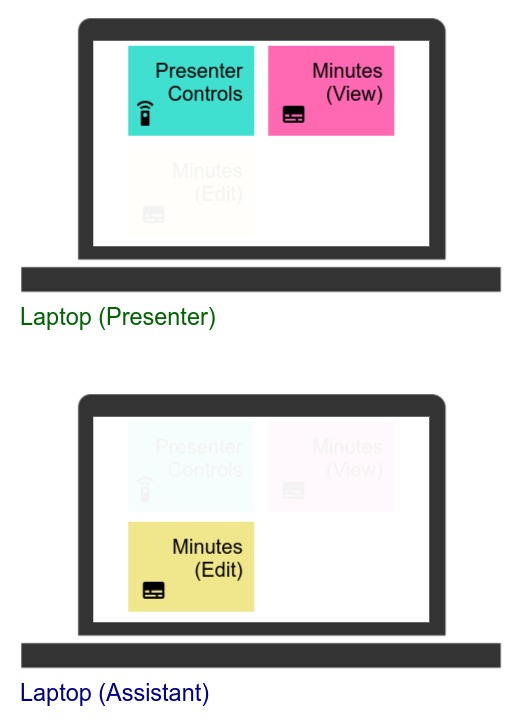}
	\vskip -3mm
	\small\raggedleft (b) Adapted
\end{minipage}
\caption{Adapting to user roles.
Giving permissions only for a subset of available elements allows for an interface which satisfies the requirements of users' roles.}
\label{fig:evaluation_roles}
\end{figure}

Figure~\ref{fig:evaluation_roles} shows that setting permission values only, already yields meaningful results. While the initial layout has no awareness of user roles (a), our algorithm correctly removes UI elements for unauthorized users (b).

\subsection{B. User Preference}
\begin{figure}
\centering
\includegraphics[width=0.8\columnwidth]{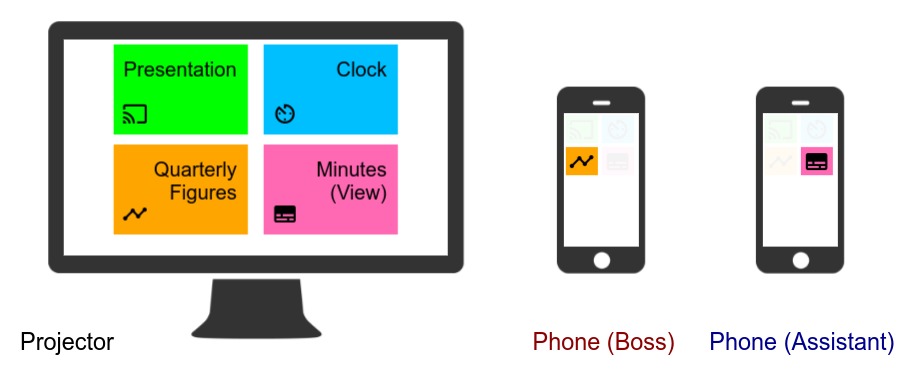}
\caption{Adapting to user preferences.
Initially, all elements appear on the projector.
Increasing the per user importance of individual elements, triggers appearance on personal devices.
}
\label{fig:evaluation_preferences}
\end{figure}

While user roles are respected via a designer-specified constraints, user preference is accounted for by the optimization objective.
This allows for a flexible balancing of preferences, which is shown further in the demo application section.
We show a simple example in Figure~\ref{fig:evaluation_preferences}.
Initially all four UI elements have the same importance values and are therefore displayed on a large shared screen with a random element assigned to personal devices.
Once the boss and assistant set higher importances for the ``Quarterly Figures'' and ``Minutes (View)'' elements, these are assigned to their personal devices.
Note that in this example, the size, input and output requirements of elements and the device characteristics are kept equal.
Examples in our demo application in the next section show more sophisticated changes in user preference.

\subsection{C. Device Compatibility}
\begin{figure}
\centering
\begin{minipage}{\columnwidth}
	\centering
	\includegraphics[width=0.8\columnwidth]{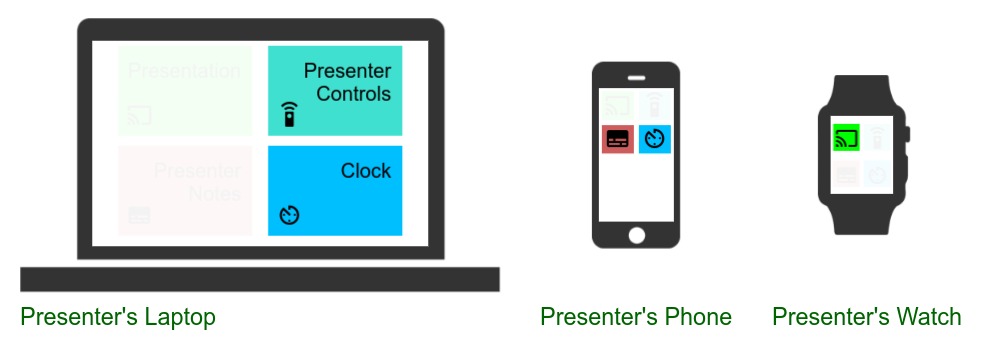}
	\vskip -0mm
	\small\raggedright (a) Without device characteristics or element requirements
\end{minipage}
\hfill\vline\hfill
\begin{minipage}{\columnwidth}
	\centering
	\includegraphics[width=0.8\columnwidth]{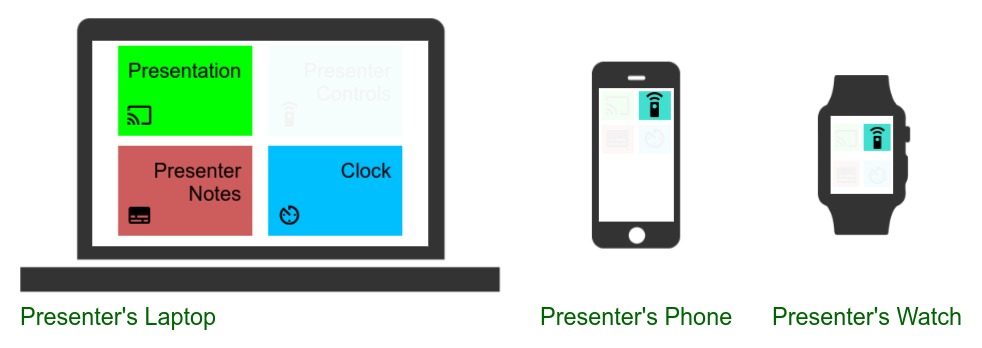}
	\vskip -0mm
	\small\raggedright (b) With sensible device characteristics and element requirements parameters
\end{minipage}
\vskip 1mm
\caption{
Element assignments become more suitable when taking in to consideration device characteristics and element requirements.
}
\label{fig:evaluation_compatibility}
\end{figure}

We attempt to assign each UI element to the most suitable devices by considering element-device compatibility.
We show an example of a single presenter with 3 devices, shown in Figure~\ref{fig:evaluation_compatibility}.
We compare (a) a case where all parameters are set to $1$ against (b) a case with sensible parameters.
Exact parameters are listed in the Appendix (Tab.~\ref{tab:evaluation_compatibility}).
Note that other input parameters are kept fixed and that all elements including the presentation slides fit onto the smartwatch's display.

Clearly a naive distribution of elements onto devices does not make sense since there is no guidance in terms of device affordances.
The ``Presenter Notes'' element is placed on a small smartphone while the ``Presentation'' element is placed on the even smaller smartwatch.
While ``Presenter Controls'' may be used on a laptop, arguably this element would be better placed on the available touchscreen device.
In contrast, by setting sensible device characteristics and element requirement parameters, we can attain a useful assignment.
While a human designer may not have duplicated the ``Presenter Controls'' over the smartphone and smartwatch and may have moved the ``Clock'' to the watch, we note that this is simply an initial assignment and can be refined quickly by tuning further input parameters such as setting the correct element size bounds and adjusting importance values. Since optimization takes only seconds this can be done interactively.

\subsection{D. Individual UI Completeness}
An important contribution of our work is a formulation that considers completeness of the final DUI.
When elements are assigned to devices without the completeness term or consideration of element utility from each user's perspective, a particular user may receive an incomplete and hence non-functional UI.
We address this by encouraging the optimizer to maximize the number of elements that a user can utilize.

\begin{figure}
\begin{minipage}[h]{\columnwidth}
	\centering
	\includegraphics[width=0.6\columnwidth]{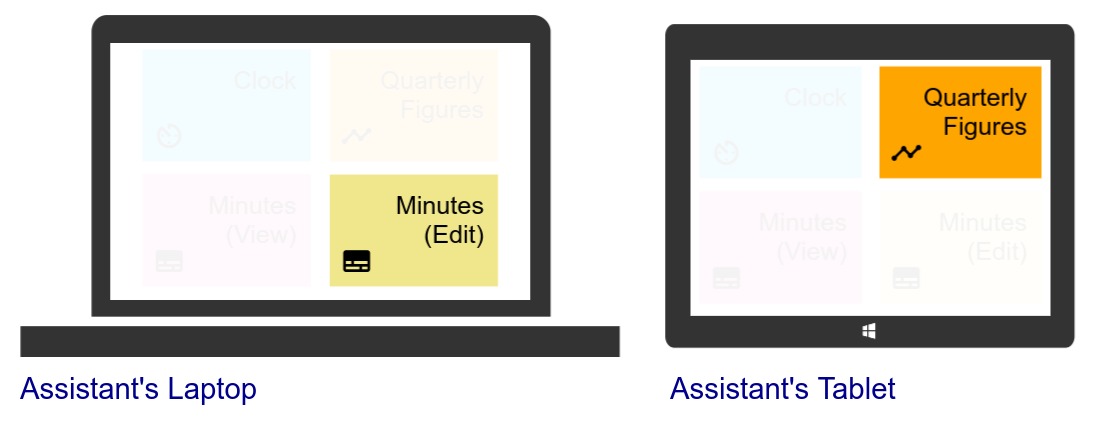}
	\raisebox{1cm}{$~\implies~$}
	\includegraphics[width=0.251\columnwidth]{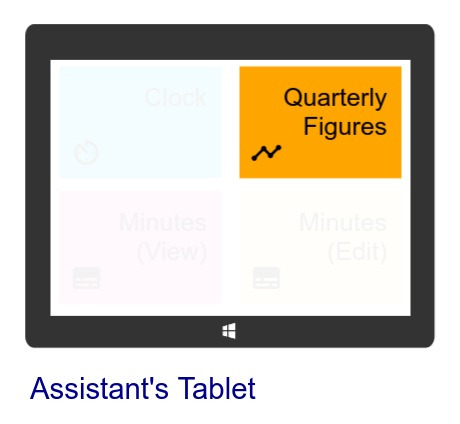}
	{\small (a) Without Completeness Term}
\end{minipage}
\begin{minipage}[h]{\columnwidth}
	\centering
	\includegraphics[width=0.6\columnwidth]{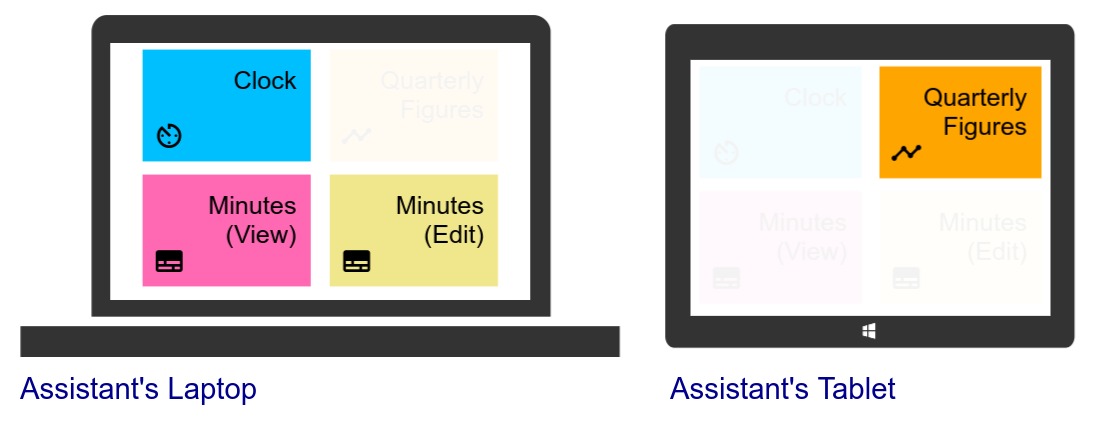}
	\raisebox{1cm}{$~\implies~$}
	\includegraphics[width=0.251\columnwidth]{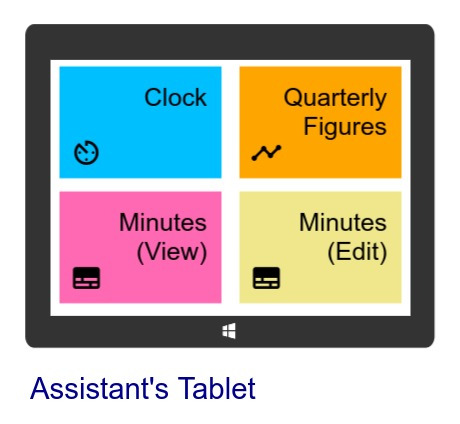}
	{\small (b) With Completeness Term}
\end{minipage}
\vskip 1mm
\caption{The completeness term ensures that the final DUI remains useful.
(a) shows the low utility of the DUI generated by the optimizer without the completeness term while,
(b) shows how all elements are available for the user when using the completeness term.}
\label{fig:evaluation_completeness}
\end{figure}

Figure~\ref{fig:evaluation_completeness} shows the effect of the completeness term.
The original UI shown in (a) is incomplete, and switching the laptop off only exaggerates this issue, where the assistant is left with a single UI element.
When adding the completeness term, the initial UI includes all available elements (b).
After switching the laptop off, the three elements previously assigned to the laptop move to the tablet and the UI remains functional.

By introducing the DUI completeness term which improves the functionality of each user's DUI, we ensure that utility is part of the optimizer's objective.
Our consideration results in usable DUIs and is a meaningful step towards optimizing for individual users in a multi-user setting.

%!TEX root = ../proceedings.tex

\section{Demo Application: Co-located Media Sharing}
\begin{figure*}
\centering
\begin{minipage}[t]{\textwidth}
	\includegraphics[width=\columnwidth]{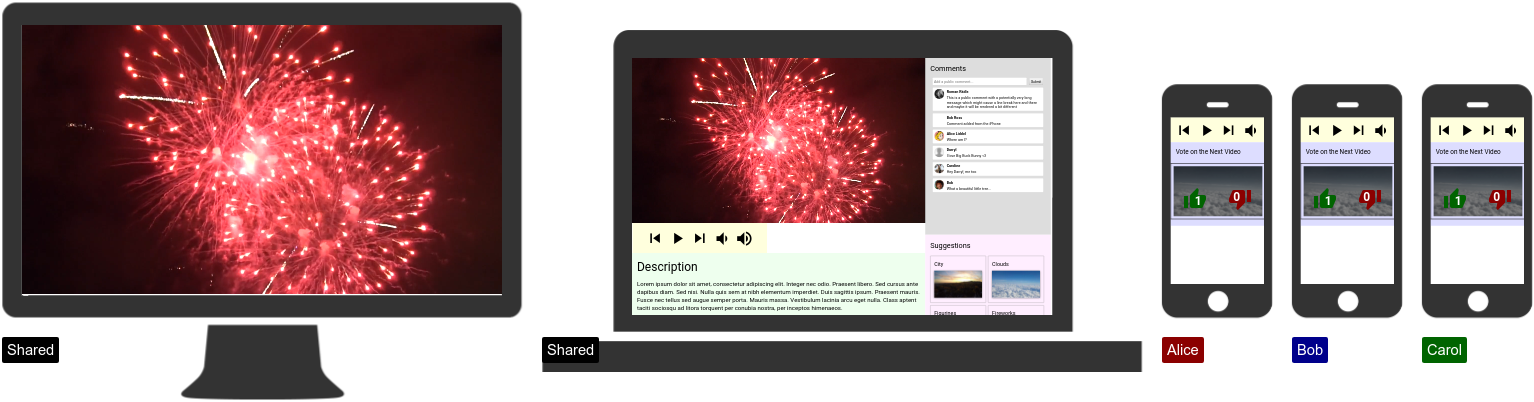}
	{\small
	(a) Initial configuration.
	It can be seen that elements respect element-device compatibilities in their assignment.}
\end{minipage}
\vskip 1mm
\begin{minipage}[b]{0.88\textwidth}
	\includegraphics[width=0.46\columnwidth]{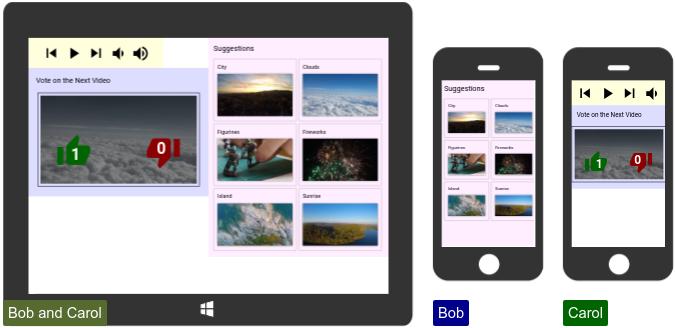}
	\hfill\raisebox{1.6cm}{$~\implies~$}\hfill
	\includegraphics[width=0.46\columnwidth]{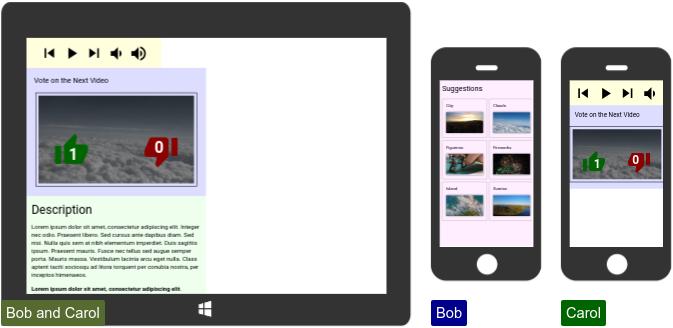}\\
	{\small
	(b) Bob and Carol's preferences can both be respected.
	On the left, only Bob's preference of the ``Suggested Videos'' element is represented.
	On the right, Carol's preference for the ``Description'' element is also addressed by placing the element on the tablet shared with Bob.}
\end{minipage}
\hfill\vrule\hfill\hskip 1mm
\begin{minipage}[b]{0.09\textwidth}
	\centering
	\includegraphics[width=\columnwidth]{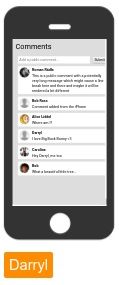}
	{(c)}
\end{minipage}
\vskip 1mm
\caption{A demonstration of our full system with optimization backend and distributed frontend.
In this example, we can see $4$ users and $7$ devices in play with three user preferences represented.
Our system quickly adapts to the changing setting with ease.
(a) and (b) are explained in their own captions and
(c) reflects Darryl's preference for reading comments.
}
\label{fig:demo}
\end{figure*}

After analyzing the individual components of our approach we now discuss a more end-to-end application that we implement using the proposed optimization approach.
In our application, we explore the task of co-located media sharing, being particularly well suited to demonstrate the capabilities to adapt to dynamic changes. This is one of the main contributions of our work and have previously not been modeled.
Our approach makes it possible to adapt to arbitrary changes in a scenario in real-time and allows a designer or even the end-users to express and apply their preferences to continuously improve the user experience.
In this application, we design our elements using responsive web design practices.
The result is a visually appealing and functional application.

We consider a scenario involving 4 users, shared devices with large displays as well as smaller private devices.
The UI consists of the following elements: video, playback controls, description, comments, and suggested videos.
We also add a collaborative component by implementing a voting module.
When a user clicks on one of the suggested videos, the video is shown on the voting element.
When all users have voted, the vote concludes and the suggested video may be played.

In our scenario we begin with 1 TV, 1 shared laptop, and 3 smartphones.
We do not illustrate all devices in the paper and refer the reader to our supplementary video for a visual demonstration of how our system handles dynamic user, device and user preference changes.

\subsection{Initial Condition}
Without any user preferences expressed, our algorithm can still produce sensible element assignments taking element size ranges, device characteristics, element requirements, and device sizes into consideration.
Figure~\ref{fig:demo}a shows the optimized assignment in the AdaM simulator UI.
It can be seen that the most visually important video element is placed on the shared large displays, while the voting controls which require touch interaction are placed on the mobile phone displays.
The comments element requires text input, and is appropriately placed on the laptop.

\subsection{Bob and Carol's preferences and shared tablet}
During the video sharing session, Bob and Carol bring out their tablet.
When Bob increases his importance value for the ``Suggestions'' element to be higher (5) than the default for everyone else (4), the element appears on the tablet.
With an even higher importance value of $8$, the suggestions appear on the phone as well, replacing the voting element (see Figure~\ref{fig:demo}b).

When Carol decides that she would like to read the description of the video, she sets an importance value that is higher than Bob's importance for the suggestions element.
She has to set a sufficiently higher value of $14$ however, to counter-act the lower compatibility between the description element and tablet.
The result of this is shown in Figure~\ref{fig:demo}b as well.
Our completeness measure ensures that both Carol and Bob can still access the important voting controls.

\subsection{Darryl joins with his own preference}
Later in the evening, Darryl joins the gathering.
He prefers to read other users' opinions, and therefore he places a high importance on the comments element.
When he sets his personal importance value for the comments element to $10$, it is placed on his personal smartphone.
He can then read and comment as he pleases.
This result is shown in Figure~\ref{fig:demo}c.

\section{Scalability}
Our algorithm is capable of adapting to changes in users, devices, and elements in real-time.
So far and for brevity we discussed only toy examples in which the run-time of the optimizer was $\approx 0.1$s.
Here we evaluate how well the algorithm scales to larger number of devices, elements and users, settings in which manual assignment would be at best tedious if not impossible.
We run our performance evaluations on a desktop PC with an Intel i7-4770 processor and $32$GB of RAM.
Gurobi $7.5$ is used to solve our optimization problem.

As a test for worst-case scenarios, we randomly generate a large number of elements, devices, and users, and record convergence time of the solver over 10 randomized runs.
For users, random per-element importance values are generated and devices are generated with width/height values between $200$px and $1200$px.
We allow all users to access all devices.
Elements are generated with minimum width/height of $100$px and maximum width/height of $1600$px with $10$px increments between randomized values.

\begin{figure}[htb]
\begin{minipage}[t]{0.49\columnwidth}
\includegraphics[width=\columnwidth]{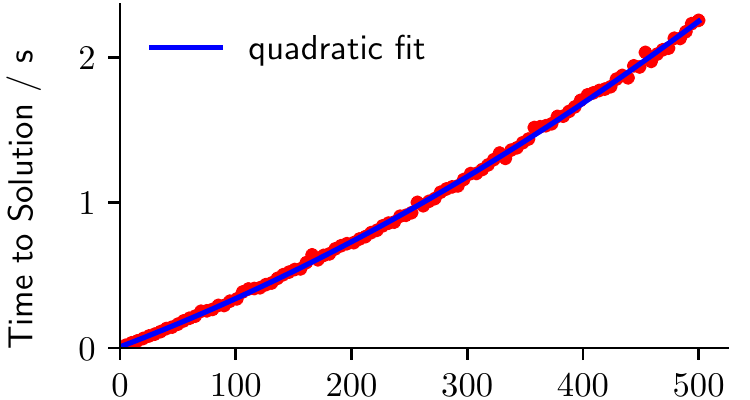}
\centering{\small(a) Elements}
\end{minipage}
\hfill
\begin{minipage}[t]{0.49\columnwidth}
\includegraphics[width=\columnwidth]{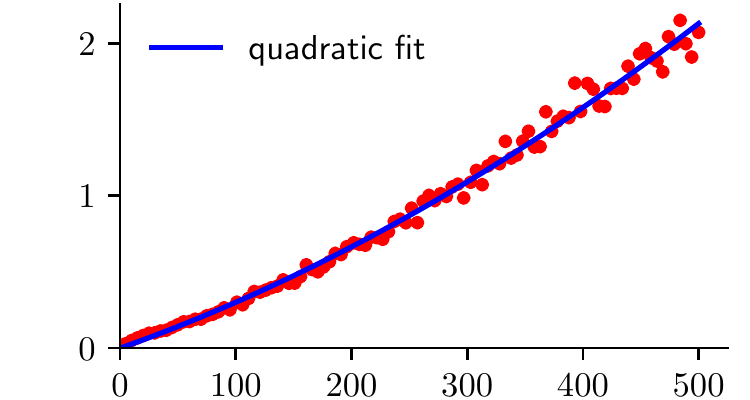}
\centering{\small(b) Devices}
\end{minipage}
\\
\begin{minipage}[t]{0.49\columnwidth}
\includegraphics[width=\columnwidth]{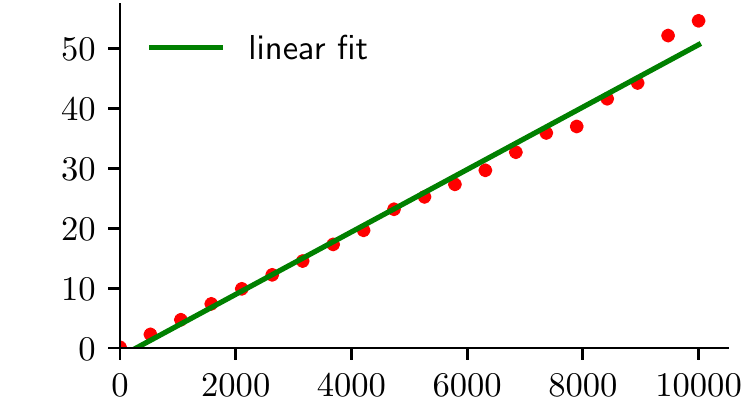}
\centering{\small(c) Users}
\end{minipage}
\hfill
\begin{minipage}[t]{0.49\columnwidth}
\includegraphics[width=\columnwidth]{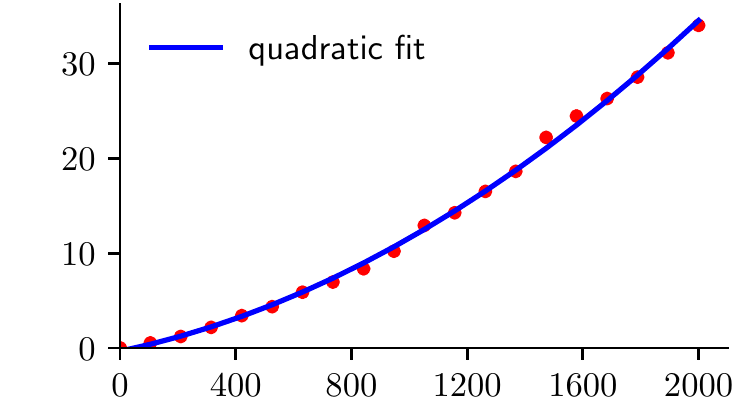}
\centering{\small(d) Users (and Devices)}
\end{minipage}
\vskip 1mm
\caption{Optimization time in seconds for varying problem sizes. In (a-c) we vary the number of elements, devices, and users independently. In (d) we vary both users and devices.}
\label{fig:scaling}
\end{figure}

Figure~\ref{fig:scaling} summarizes the results.
In (a), the input data consists of $20$ devices and $10$ users with an increasing number of elements.
In (b), we input $20$ elements and $10$ users with an increasing number of devices.
In (c), there are $50$ devices, $20$ elements, and up to $10^4$ users to show an extreme scenario.
All users have randomized personal preferences.
To consider a more realistic case, we fix the number of elements to $20$ and vary both users and devices in (d).
There are $2$ personal devices per user and $1$ publicly shared device per 5 users.

Our algorithm can solve a scenario with $100$ users and $220$ devices in $\approx 1$ second, allowing for the design of large-scale real-time adaptive systems.
This speed allows for a real-time exploration of DUI configurations where a designer can determine parameters suitable to a task based on instant feedback.

%!TEX root = ../proceedings.tex
\section{User Study}

We assessed the approach by asking experimental participants to design a DUI using either pen and paper or AdaM.
Our goal was to understand whether our approach is easy to understand, and to see if we can observe improvements in the design process in terms of performance and experience.

\subsection{Method}

\emph{Participants:} Six participants (3 female, 3 male) were recruited from our institution (students and staff). The average age was 26 (SD = 1.6, aged 24 to 27). Two participants were researchers in the area of web engineering with one of them in particular researching DUIs. Three other participants stated to have web development experience.

\emph{Tasks:} The study comprised of two tasks centered around a meeting scenario:
\begin{inparaenum}[1)]
	\item Participants were asked to assign UI-elements to devices to reflect the role and preference of users as specified in the scenario (T1).
	\item In the second task, some devices were switched on/off and content preferences were changed. Participants were asked to adapt the previous assignment accordingly (T2).
\end{inparaenum}

\begin{figure}[tbh]
	\centering
    \includegraphics[width=\linewidth]{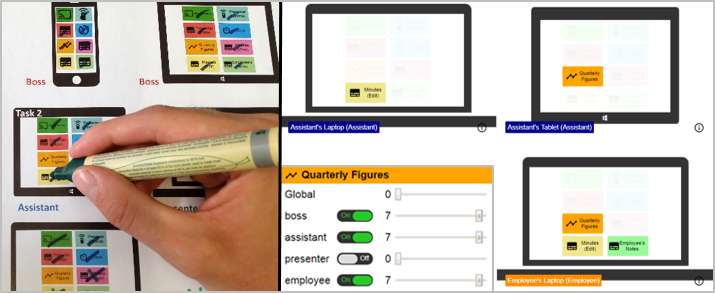}
	\caption{Conditions of study (left: \emph{pen\&paper}, right: \emph{AdaM}).}
	\label{fig:conditions}
\end{figure}

\emph{Experimental design:}
We tested two conditions.
In the first condition (\emph{pen\&paper}), participants crossed out elements which did not match the given scenario on a large sheet of paper showing all devices of all roles (see Figure \ref{fig:conditions}, left).
In the second condition (\emph{AdaM}), participants used sliders to specify element importance according to scenario descriptions.
An additional UI displayed an overview of devices and assigned elements (see figure \ref{fig:conditions}, right).
We used a within-subjects design and counterbalanced the order of presentation.

\emph{Procedure:} In the beginning, participants were introduced into \emph{pen\&paper} and \emph{AdaM} and were provided time to practice using the tool.
After that participants solved T1 and T2 in the respective conditions.
Tasks were completed when participants reported to be satisfied with the element to device assignment.
For each task and condition, participants completed the NASA-TLX and a questionnaire on satisfaction with results.
At the end an exit interview was conducted.
A session took on average 60 mins.

\subsection{Results}
In terms of perceived scenario, results satisfaction, number of scenario violations and perceived task load, the mean of responses of both conditions were within standard deviation.
However, task execution time (TET) was lower for \emph{pen\&paper} compared to \emph{AdaM}\footnote{For a summary of quantitative results see Table~\ref{tab:results} in the appendix.}, which indicates that the design task may not have been sufficiently difficult.
This highlights the challenge of performing a fair comparison between automatic and manual designs from the designers' perspective, where the task cannot be so difficult to be deemed unfair.

Analyzing the answers of the interviews, three participants valued \emph{AdaM}'s capability to
adapt in real-time to changing device configurations.
In fact, one participant even exclaimed ``perfect!'' after switching on a mobile phone and realizing that the automatically assigned UI elements satisfy the scenario without any further adjustment.
This advantage is also evident when looking at the differences of quantitative results between the assignment task T1 and the adaption task T2.
In between tasks, the average TET improved by 103 sec with \emph{AdaM} compared to only 14 sec in \emph{pen\&paper} and task load improved by 14.6 with \emph{AdaM} compared to 6.2 in \emph{pen\&paper}.

Another property of \emph{AdaM} that was perceived as a ``powerful'' advantage over the manual
approach (5 out of 6 participants) was the possibility to specify ``global rules'' (so named by a participant). They liked the fact that instead of assigning elements on a device level, they could specify the preference of a person and let the optimizer distributes elements over her devices.
Participants commented on this capability saying “not white and black listing per device, but you
specify importance per role” or ``when I specify the importance I do not
need to think about devices''.

Nevertheless, the same participants mentioned that the main drawback of \emph{AdaM} was less control in terms of specifying distinct element to device assignments. They struggled with finding a balance between different slider values such that the optimizer's element-to-device-assignment matches their intention. One participant summarized that problem with: ``I was able to satisfy the scenario, [but] it was difficult with the optimizer to go beyond''. A solution for this problem is to allow the specification of element-to-device assignments as hard constraints (see paragraph \emph{User-defined Element Assignment}).

Another difficulty participants had was to understand the expected outcome of a slider change (``what does it translate to when I set a slider to 15?'').
Due to the non-linear nature of our formulation the outcome of the optimizer is hard to predict and thus how sliders need to be adjusted.

%!TEX root = ../proceedings.tex

\section{Limitations and Future Work}
In this paper, we laid the foundations for future work but it is not without limitations.
User study participants in particular had difficulty predicting the optimizer's output (i.e., when the size bounds of the video element changes, how does the output change?),
while the large number of input parameters and the difficulty of determining the best parameters caused some difficulty in implementing the demo application.
These issues could be addressed by:
\begin{inparaenum}[(a)]
\item producing a rigorous DUI test framework based on empirical observations (to allow for an improved objective function formulation),
\item reducing the number of input parameters (e.g., by defining a mapping from real-world device characteristics to $\mathbf{u}_d$ or using user interaction logs for determining $i_{eu}$),
and
\item improving the DUI design-space exploration experience for designers (e.g., facilitating easy specification of scenarios and automated mockup of heterogeneous set of devices associated with users).
\end{inparaenum}

A further limitation is in our evaluations.
While our user study serves its purpose of confirming the general idea of our approach, low participant numbers and the simulated design task cause us to hesitate in forming generalized conclusions.
Nevertheless, we have confidence in our approach as it was designed to be general and user-centred, with basic principles in mind.
Thus, we believe that AdaM can be effective in real world settings and aim to conduct an in-depth analysis in the future to verify our thoughts.

Further extensions to improve user experience in AdaM-based DUIs could include:
\begin{inparaenum}[(1)]
	\item consideration of user proximity and attention for $a_{ud}$,
	\item automatic determination of element-device compatibility parameters $\mathbf{u}_d$ and $\mathbf{v}_e$ based on the affordances of devices and composition of elements,
	and
	\item continuous adaptation to users' changing preferences through analysis of interaction logs and visual attention tracking.
\end{inparaenum}

\section{Conclusion}
In this paper we have demonstrated a scalable approach to the automatic assignment of UI elements to users and devices in cross-device user interfaces, during multi-user events.
By posing this problem as an assignment problem, we were able to create an algorithm which adapts to dynamic changes due to altering configurations of users, their roles, their preferences and access rights, as well as advertised device capabilities.

Underpinning AdaM, is a MILP solver which given an objective function decides the assignment of elements to multiple devices and users.
Measures for both quality, completeness along with constraints, help to guide the optimization toward satisfactory solutions, which are represented by suitable assignments of UI elements.
Following this, the layout problem is performed by responsive design practices common in web design, as shown in our application scenarios.

The AdaM application platform itself is web-based and enables collaborative prototyping and rapid iterations of AdaM applications.
In addition, our simulator environment allows us to instantiate a wide range of simulated devices.
We report on scenarios with up to 1000 users and 2200 devices along with a user study involving six participants, who are asked to assign and adapt UI-element configurations.
Our qualitative results indicate that AdaM can reduce both designer and user effort in attaining ideal DUI configurations.
The results are promising and suggest further exploration is warranted into the automatic UI element assignment approach introduced here.

The mathematical formulation introduced here may be extended to incorporate other issues present in collaborative multi-user interfaces including, extended device parameterization, social acceptability factors, user attention, proxemic dimensions, display switching, display contiguity, field of view, spatio-temporal interaction flow, inter-device consistency, sequential and parallel device use along with synchronous and asynchronous device arrangements.

\section{Acknowledgments}
We thank the ACM SIGCHI Summer School on Computational Interaction 2017 for bringing the authors together along with our study participants and the reviewers of this work.

This work was supported in part by ERC Grants OPTINT (StG-2016-717054) and
Computed (StG-2014-637991),
SNF Grant (200021L\_153644),
the Aarhus University Research Foundation,
the Innovation Fund Denmark (CIBIS 1311-00001B),
and the Scottish Informatics and Computer Science Alliance (SICSA).

% Balancing columns in a ref list is a bit of a pain because you
% either use a hack like flushend or balance, or manually insert
% a column break.  http://www.tex.ac.uk/cgi-bin/texfaq2html?label=balance
% multicols doesn't work because we're already in two-column mode,
% and flushend isn't awesome, so I choose balance.  See this
% for more info: http://cs.brown.edu/system/software/latex/doc/balance.pdf
%
% Note that in a perfect world balance wants to be in the first
% column of the last page.
%
% If balance doesn't work for you, you can remove that and
% hard-code a column break into the bbl file right before you
% submit:
%
% http://stackoverflow.com/questions/2149854/how-to-manually-equalize-columns-
% in-an-ieee-paper-if-using-bibtex
%
% Or, just remove \balance and give up on balancing the last page.
%
%\balance{}
\clearpage

% REFERENCES FORMAT
% References must be the same font size as other body text.
\bibliographystyle{SIGCHI-Reference-Format}
\bibliography{references}

\clearpage
\section{Appendix}
\subsection{Device Capability Study Parameters}
\begin{table}[!h]
\centering

\begin{minipage}[b]{\columnwidth}
\centering
\begin{tabular}{|l|ccc|}
\hline
& Laptop & Smartphone & Smartwatch \\
\hline
Visual Quality & 3 & 1 & 1 \\
Text Input     & 5 & 3 & 0 \\
Mouse Pointing & 3 & 0 & 0 \\
Touch Pointing & 0 & 4 & 2 \\
\hline
\end{tabular}
\vskip 2mm
{(a) Device Characteristics}
\end{minipage}
\vskip 3mm

\begin{minipage}[b]{\columnwidth}
\centering
\begin{tabular}{|m{1.5cm}|>{\centering}m{1.45cm}>{\centering}m{1.3cm}>{\centering}m{1.2cm}>{\centering}m{0.8cm}|}
\hhline{-----}
& \cellcolor{Presentation} Presenta-tion
& \cellcolor{PresenterControls} Presenter Controls
& \cellcolor{PresenterNotes} Presenter Notes
& \cellcolor{Clock} Clock \tabularnewline
\hline
Visual Quality & 5 & 0 & 3 & 2 \tabularnewline
Text Input     & 0 & 0 & 1 & 0 \tabularnewline
Mouse Pointing & 0 & 3 & 1 & 0 \tabularnewline
Touch Pointing & 0 & 5 & 1 & 0 \tabularnewline
\hline
\end{tabular}
\vskip 2mm
{(b) Element Requirements}
\end{minipage}

\caption{Our device characteristics and element requirements parameters for Fig.~\ref{fig:evaluation_compatibility} (b).}
\label{tab:evaluation_compatibility}
\end{table}

\subsection{User Study Quantitative Figures}
In the study, we asked participants whether the assignment of elements they produced in a condition satisfies the scenario (on a scale ranging from 1 (not at all) to 7 (completely)) and how satisfied they were with the assignment they specified (ranging from 1 (not satisfied) to 7 (very satisfied)). The results of these questions as well as the task execution time for both conditions and tasks can be seen in table \ref{tab:results}. Furthermore, we asked participants to fill out the Nasa-TLX questionnaire and calculated how often the designed element-to-device assignments of participants violated the given scenario for both conditions and tasks. Results are again shown in table \ref{tab:results}.
\begin{table}[tbh]
	\centering
	\setlength{\tabcolsep}{3pt}
	\begin{tabular}[c]{|l|l||l|l|}
		\hline
		Task&Measure&\emph{pen\&paper}&\emph{AdaM}\\
		\hline
		T1&Exec. time ($s$)&313$\pm$90&399$\pm$119\\
		%&NASA TLX&36.9$\pm$17.4&41.7$\pm$10.8\\
		&Scenario satisfaction&5.2$\pm$2.1&5.7$\pm$1.6\\
        &Result satisfaction&4.8$\pm$1.9&4.8$\pm$1.6\\
        &Scenario violations&1$\pm$1.5&1.3$\pm$1.2\\
        &Task load&36.9$\pm$17.3&41.7$\pm$10.8\\
        		\hline
        T2&Exec. time ($s$)&259$\pm$141&296$\pm$112\\
		%&NASA TLX&30.7$\pm$16.9&27.1$\pm$13.3\\
		&Scenario satisfaction&6$\pm$0.6&5.7$\pm$0.8\\
        &Result satisfaction&5.5$\pm$1.8&5.3$\pm$1.0\\
        &Scenario violations&1.2$\pm$1.0&1.2$\pm$1.3\\
        &Task load&30.7$\pm$16.9&27.1$\pm$13.3\\
        \hline
	\end{tabular}
	\caption{Mean and SD of quantitative measures for T1 and T2 per condition.}
	\label{tab:results}
\end{table}

\end{document}